\documentclass[twocolumn]{aastex631}
\turnoffedit

\newcommand{\comment}[1]{}

\usepackage{amsmath, amsbsy, soul}

\graphicspath{{./}{figures/}}

\begin{document}

\title{Magnetic fields in the Southern Coalsack and beyond}

\author[0000-0003-0400-8846]{M.J.F. Versteeg}
\affiliation{Department of Astrophysics/IMAPP, Radboud University, PO Box 9010,
6500 GL Nijmegen, The Netherlands}

\author[0000-0001-5016-5645]{Y. Angarita}
\affiliation{Department of Astrophysics/IMAPP, Radboud University, PO Box 9010,
6500 GL Nijmegen, The Netherlands}

\author[0000-0002-1580-0583]{A.M. Magalh{\~a}es}
\affiliation{Depto. de Astronomia, IAG, Universidade de São Paulo, Brazil}

\author[0000-0002-5288-312X]{M. Haverkorn}
\affiliation{Department of Astrophysics/IMAPP, Radboud University, PO Box 9010,
6500 GL Nijmegen, The Netherlands}

\author[0000-0002-9459-043X]{C.V. Rodrigues}
\affiliation{Divis\~ao de Astrof\'\i sica,
Instituto Nacional de Pesquisas Espaciais (INPE/MCTI),
Av. dos Astronautas, 1758,
S\~ao Jos\'e dos Campos, SP, Brazil}

\author[0000-0001-6880-4468]{R. Santos-Lima}
\affiliation{Depto. de Astronomia, IAG, Universidade de São Paulo, Brazil}

\author[0000-0001-6099-9539]{Koji S. Kawabata}
\affiliation{Hiroshima Astrophysical Science Center, Hiroshima University,
Kagamiyama, Higashi-Hiroshima, Hiroshima, 739-8526, Japan}



\begin{abstract}

Starlight polarimetry, when combined with accurate distance measurements, allows for exploration of the three-dimensional structure of local magnetic fields in great detail. We present optical polarimetric observations of stars in and close to the Southern Coalsack, taken from the Interstellar Polarization Survey (IPS). Located in five fields of view approximately 0.3\degr \space by 0.3\degr \space in size, these data represent the highest density of optical polarimetric observations in the Southern Coalsack to date. Using these data, combined with accurate distances and extinctions based on Gaia data, we are able to characterize the magnetic field of the Coalsack and disentangle contributions to the polarization caused by the Southern Coalsack and a background structure. For the Southern Coalsack, we find an average magnetic field orientation of $\theta\sim 75\degr$ with respect to the Galactic north pole and an average plane-of-sky magnetic field strength of approximately $B_{POS}=10$ $\mu G$, using the Davis-Chandrasekhar-Fermi (DCF) method. These values are in agreement with some earlier estimates of the Coalsack's magnetic field. In order to study the distant structure, we introduce a simple method to separate and isolate the polarization of distant stars from foreground contribution. For the distant structure, which we estimate to be located at a distance of approximately 1.3-1.5 kpc, we find an average magnetic field orientation of $\theta\sim100\degr$ and we estimate a field strength of $B_{POS}\sim10 \ \mu G$, although this will remain highly uncertain until the precise nature of the distant structure can be uncovered.

\end{abstract}



\section{Introduction} \label{sec:intro}

Optical starlight polarization is a powerful tracer of interstellar material and magnetic fields. \citep{ hall1949Sci...109..166H, hiltner1949polarization}. Asymmetric dust grains in the interstellar medium (ISM) align themselves with the Galactic Magnetic Field (GMF) \citep[see e.g.][]{lazarian2007tracingbfields} and subsequently preferentially absorb a fraction of the radiation polarized perpendicularly to the magnetic field lines, see e.g. \citet[][]{clarke2010stellarpolarimetry, whitett2022dustgalacticenvironment}. Thus, stars that emit unpolarized light may show partial polarization when shining through the dusty ISM. The polarization angle corresponds to the local orientation of the plane-of-sky component of the magnetic field and the degree of polarization may be indicative of the dust conditions along the line of sight (LOS), as well as the magnetic field orientation. In addition, the dispersion of the polarization angles can be used to determine the strength of the GMF, using for example the Davis-Chandrasekhar-Fermi method \citep{davis1951strength, chandrasekhar1953magnetic}, hereafter DCF. For a dominating regular magnetic field with an additional, but relatively small turbulent component, the polarization angles would follow a roughly Gaussian distribution, centered around the local orientation of the plane-of-sky component of the magnetic field. Large-scale deviations from a simple Gaussian distribution may be indicative of, for example, changes in a large-scale magnetic field direction along the LOS or the existence of intervening structures such as clouds. 

The Southern Coalsack \citep[see also][for an overview]{nyman2008southerncoalsack} is a nearby large dark nebula located in the Galactic plane, centered at approximately l=303\degr, b=0\degr \space with an on-sky size of about 10 degrees, see Figure \ref{fig:fieldloc_on_co}. It is located at a distance of $d\sim200 \textrm{ pc}$ (see e.g. \citealt[][]{franco1989coalsack}, who finds a distance of $d=180\pm26 \textrm{ pc}$, \citealt[][]{seidensticker1989distancestructurecoalsack} who identify two overlapping clouds with the closest at a distance of $d=188\pm4.1 \textrm{ pc}$ and \citealt[][]{dharmawardena3dmolclouds2023}, who place the Coalsack at a distance of approximately $d\sim150\textrm{ pc}$). The Coalsack has been studied intensively at various wavelengths, in CO by \citet{nyman1989CO_dht35} and \citet{kato1999nantencoalsack}, revealing a large and complex fragmented structure; in X-rays and FUV by \citet{andersson2004hotenvelope} showing a large hot envelope surrounding the Coalsack; and in $uvby$ and $H_{\beta}$ by \citet{franco1989coalsack}, who studied the dark cloud complex as well as the regions immediately in front of and behind the Coalsack. \citet{mcclure2001hishells} have observed HI in the direction of the Coalsack, revealing a large HI shell located behind the Southern Coalsack. Recent dust maps \citep[][]{lallement20223dinterstellardust, vergely20223dextinction, edenhofer20233ddustmap} show that the Coalsack is likely attached to the wall of the Local Bubble (LB). Furthermore, the Coalsack has been studied polarimetrically as well by \citet{jones1984innercore} in the near-infrared (NIR), \citet{andersson2005highsampling}]in the optical,  and by \citet{lada2004dustyring}, who base their calculations on observations from \citet{jones1984innercore}. The latter two studies lead to estimates of the magnetic field strength ranging from 23 to 93 $\mu$G for the component perpendicular to the LOS. The plane-of-sky orientation of the magnetic field was found to be between $\theta\sim 60\degr$ and $\theta\sim 80\degr$ \citep[see][ Table 2]{andersson2005highsampling} and \citet{jones1984innercore}, their Figure 4. Note that, in light of recent evidence that the Coalsack and LB wall are attached, it is likely that unless explicitly accounted for, these estimates of magnetic field properties for the Southern Coalsack also contain contributions from the LB wall. The angles mentioned above and throughout the rest of the paper will be given in the Galactic coordinate system, increasing from North through East.

Using high spatial density observations from the Interstellar Polarization Survey, General ISM (IPS-GI) catalog (Magalh{\~a}es et al. in prep, \citealt[][]{versteeg2023ipsII}) in combination with reliable distance measurements from \citet[][]{anders2022photo} based on Gaia EDR3 \citep[][]{gaia2021edr3}, we are able to analyze the magnetic field properties of the Southern Coalsack in great detail.

The high density of stars allows us to distinguish and separate different polarizing components or layers along the LOS and determine their individual magnetic field orientations and strengths. However, because the polarized signal comprises contributions from all structures along the LOS, the influence of foreground structures must be removed to be able to study any distant structures in isolation.

In this paper we present the highest density of stellar optical polarimetric observations in the Southern Coalsack yet, in five fields of view toward the Southern Coalsack, showing in great detail the orientation as well as the strength of the magnetic fields in the Southern Coalsack cloud complex. Furthermore, we present a simple method which can be used to isolate the background signal from foreground contamination, allowing for studies of the magnetic fields behind the Coalsack. The paper is organized as follows. We describe the polarimetric observations in Section \ref{sec:obs}. We outline our methods in Section \ref{sec:mtds}. We then apply these methods and present our results on a field-by-field basis in Section \ref{sec:res}. We discuss our results in Section \ref{sec:disc} and summarize and conclude our findings in Section \ref{sec:con}.

\section{Observations} \label{sec:obs}
\begin{figure*}[ht!]
\includegraphics[width=\textwidth]{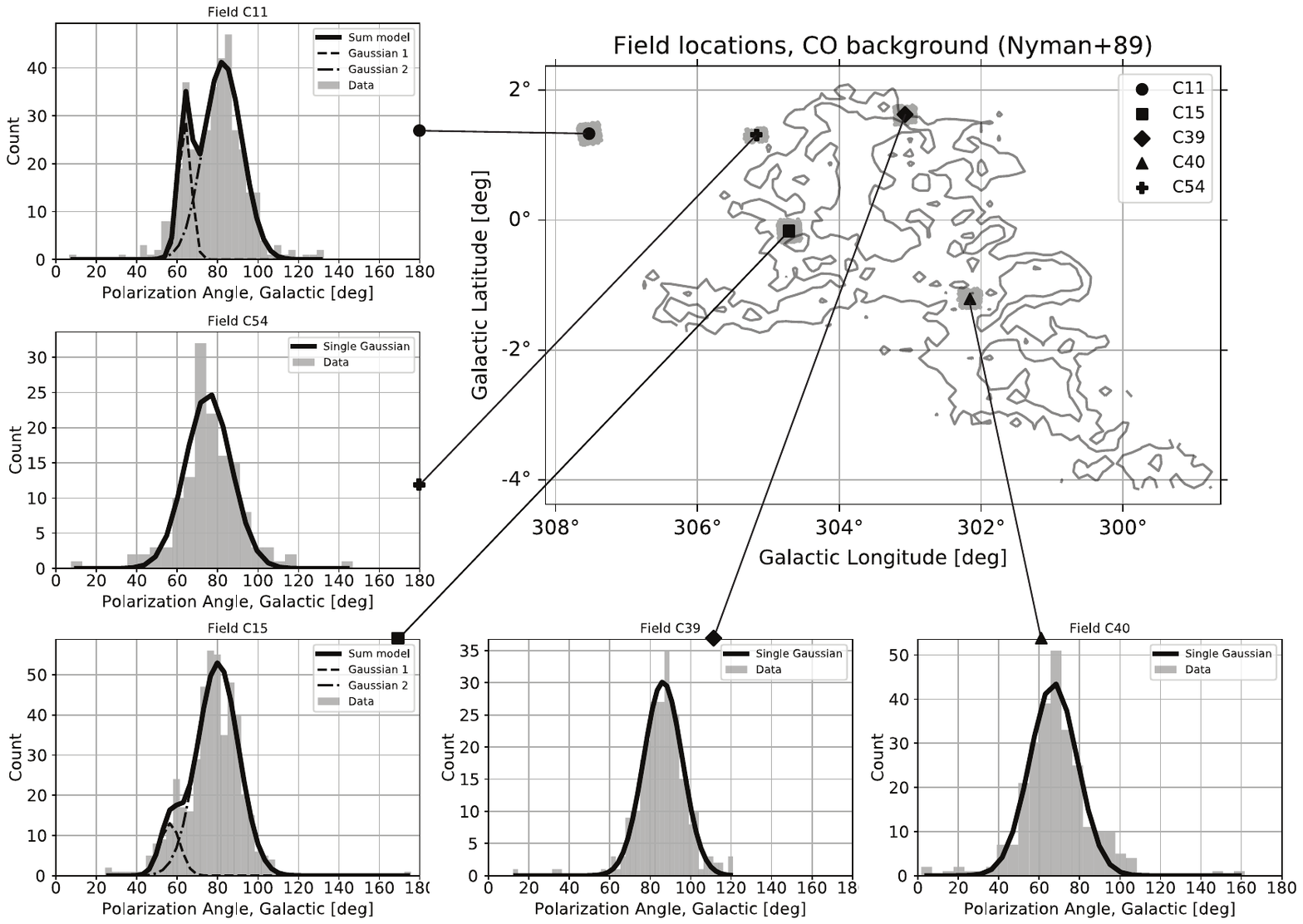}
\caption{Top right: locations of the five fields analyzed in this work. Gray rectangles represent the data, black markers indicate mean position of all stars in the field. Background contours show the total intensity in CO \citep{nyman1989CO_dht35} at the $2 \textrm{ K km s}^{-1}$ and $6 \textrm{ K km s}^{-1}$ levels. The surrounding histograms show the polarization angle distributions for each of the fields including a Gaussian fit to the distribution (see also Section \ref{sec:mtds.partitions}). Field C11, indicated with a circular marker, lies outside the area observed by \citet{nyman1989CO_dht35}. \label{fig:fieldloc_on_co}}
\end{figure*}

In this paper, we have used polarimetric observations from IPS-GI (\citealt{magalhaes2005southern, versteeg2023ipsII}; Magalh{\~a}es et al. in prep). This catalog contains linear polarimetric measurements of starlight polarization in the V-band of over 41000 stars, located in discrete fields in or close to the Galactic plane. Each field is typically 0.3\degr\ by 0.3\degr\ on the sky and contains on average 1000 stars. We have identified five fields located close to or in the Southern Coalsack. These are fields C11, C15, C39, C40, and C54, following \citet{versteeg2023ipsII} nomenclature. The positions of the fields relative to the CO emission from the Coalsack \citep{nyman1989CO_dht35} are shown in the top right of Figure \ref{fig:fieldloc_on_co}. 

In order to add distance information, we have cross-matched the IPS data to two auxiliary catalogs. Firstly, we match IPS data to the GAIA EDR3 \citep{gaia2021edr3} catalog using criteria based on position and magnitude. We find matches within a 3\arcsec \ position (corresponding to $1.5\sigma$ in the IPS position uncertainty) and 2 mag brightness (to account for the differences in IPS' V-band and GAIA's G-band magnitudes). After matching, we use GAIA's source ID to find matches with the StarHorse2021 catalog \citep{anders2022photo} which provides distances and other stellar parameters. To ensure that the analysis is applied only to the highest quality data, we have applied six quality filters, which are explained in more detail in \citet[][]{versteeg2023ipsII}.

\begin{enumerate}
    \item $p/dp > 5$, where \emph{p} is the degree of linear polarization ($p=\sqrt{q^2 + u^2}$) and \emph{dp} is its associated error; q and u are the normalized linear Stokes parameters, which ensures that only high-quality polarimetric observations are taken into consideration,
    \item $dp < 0.8\%$, which removes sources with spurious polarimetric errors,
    \item $fidelity > 0.5$, which is a quality filter defined by \citet{rybizki2022classifier} and removes spurious results from GAIA fitting routines,
    \item $sh_{outflag} = 0$, which removes spurious results from the output of the StarHorse tool used by \citet{anders2022photo}, 
    \item $C^{*}/\sigma_{C^{*}} < 5$, where $C^{*}$ is the corrected color excess and $\sigma_{C^{*}}$ the 1$\sigma$ scatter. This filter excludes sources with spurious color excess \citep[see also][equation 18 therein]{riello2021photometric},
    \item Only high-certainty matches, removing results for which the cross-matching between the IPS and GAIA EDR3 catalogs did not result in a unique match.
\end{enumerate}

Having applied the six quality filters, we retain 1543 stars divided over five fields, listed in Table \ref{tab:mastertable}.

\begin{deluxetable*}{cccccccccccc}

\tablecaption{Coalsack Observations and Data \label{tab:mastertable}}
\tablewidth{0pt}
\tabletypesize{\footnotesize}
\rotate
\tablehead{
\colhead{Field} & \colhead{Comp.} & \colhead{N. *} & \colhead{$\langle$Dist.$\rangle$}& \colhead{$\langle\textrm{A}_V\rangle$} & \colhead{$\langle$p$\rangle$} & \colhead{$\langle\theta\rangle$} & \colhead{$\sigma_{\theta}$\tablenotemark{a}} & \colhead{$\sigma_{v}$\tablenotemark{d}} & \colhead{$\rho$\tablenotemark{d}} & \colhead{$\textrm{B}_\perp$ (DCF)} & \colhead{$\textrm{B}_\perp$ (ST)} \\
\colhead{} & \colhead{} & \colhead{} & \colhead{kpc} & \colhead{mag} & \colhead{\%} & \colhead{deg} & \colhead{deg}  & \colhead{$\textrm{km s}^{-1}$} & \colhead{$10^{-22}\textrm{ g cm}^{-3}$} & \colhead{$\mu\textrm{G}$} & \colhead{$\mu\textrm{G}$}
}
\decimalcolnumbers

\startdata
~   & N        & 88  & $0.9^{+0.004}_{-0.004}$ & $0.7^{+0.01}_{-0.01}$ & $1.4 \pm 0.1$ & $64.2 \pm 1.4$  & $4.3 \pm 0.4$  &  &  &  &  \\ 
C11 & D        & 193 & $2.0^{+0.01}_{-0.01}$   & $1.4^{+0.02}_{-0.02}$ & $1.8 \pm 0.1$ & $83.8 \pm 1.6$  & ~ & ~ & ~ &~ & ~ \\ 
~   & D$_{iso.}$ & 193 & $2.0^{+0.01}_{-0.01}$   & $1.4^{+0.02}_{-0.02}$ & $1.1 \pm 0.1$ & $110.8 \pm 2.3$ & $6.8 \pm 0.7$\tablenotemark{b} & 7.5 & 3.68E-2 & $\sim 20$ & $\sim 10$  \\ 
~   & Neither     & 91  & & & ~             & ~               & ~ & ~ & ~ & ~ & ~ \\ \hline
~   & N        & 194 & $1.3^{+0.01}_{-0.01}$   & $1.5^{+0.02}_{-0.02}$ & $1.6 \pm 0.1$ & $70.2 \pm 1.5$  & $12.1 \pm 0.8$ & $0.8 \pm 0.03$ & $3.1 \pm 0.9$ & $12.5 \pm 2.1$\tablenotemark{c} & $8.1 \pm 2.0$ \\ 
C15 & D        & 267 & $2.4^{+0.02}_{-0.02}$   & $3.2^{+0.02}_{-0.02}$ & $3.6 \pm 0.1$ & $84.8 \pm 1.2$  & ~ & ~ & ~ & ~ & ~ \\ 
~   & D$_{iso.}$ & 267 & $2.4^{+0.02}_{-0.02}$   & $3.2^{+0.02}_{-0.02}$ & $1.7 \pm 0.1$ & $98.2 \pm 2.2$ & $15.0 \pm 2.5$\tablenotemark{b} & 7.5 & 3.68E-2 & $\sim 10$ & $\sim 5$ \\ \hline
C39 & Full      & 309 & $1.7^{+0.01}_{-0.01}$   & $2.0^{+0.02}_{-0.02}$ & $2.1 \pm 0.1$ & $83.2 \pm 2.0$  & $9.5\pm 0.3$  & $0.7 \pm 0.07$ & $1.2 \pm 0.4$ & $8.1 \pm 1.5$ & $4.6 \pm 1.5$ \\ \hline
C40 & Full      & 264 & $2.0^{+0.02}_{-0.02}$   & $2.3^{+0.02}_{-0.02}$ & $2.0 \pm 0.2$ & $64.5 \pm 2.4$  & $10.8 \pm 0.6$ & $0.8 \pm 0.05$ & $2.1 \pm 0.6$ & $10.8 \pm 1.9$ & $6.6 \pm 1.8$ \\ \hline
C54 & Full      & 137 & $1.4^{+0.01}_{-0.01}$ & $1.3^{+0.02}_{-0.02}$ & $1.5 \pm 0.1$ & $75.5 \pm 2.0$  & $12.0 \pm 0.6$  &  & $<0.9 \pm 0.3$\tablenotemark{e} &  &  \\ \hline
\enddata

\tablecomments{For each field: its field number (1); description of components along the LOS (2), where N is Nearby, D is Distant, D$_{iso}$ is the isolated Distant component and Neither is the data not assigned to a data cluster (Section \ref{sec:res.c11}); the number of stars in each component (3); the mean distance (4) and V-band extinction (5); the weighted average degree of polarization (6); the weighted average polarization angle of that component (7); the polarization angle dispersion (8); the LOS velocity dispersion (9); the local density (10) and the resulting magnetic field strength from the DCF and ST methods, (11) and (12), respectively.}
\tablenotetext{a}{The polarization angle dispersion has been corrected with the mean error in the polarization angle as in Equation \ref{eq:polangdisp}.}
\tablenotetext{b}{Isolated distant $\sigma_{\theta}$ was corrected for the contribution of the nearby component angle dispersion using Equation \ref{eq:poldisp_corr}.}
\tablenotetext{c}{Using a structure function of polarization angles, see also Section \ref{sec:disc.sf_YA}, we find a magnetic field strength $B = 31.2\pm4.7 \ \mu$G.}
\tablenotemark{d}{Based on CO data from \citet{nyman1989CO_dht35}}
\tablenotemark{e}{The density is an upper limit, see Section \ref{sec:res.c54}.}
\end{deluxetable*}


\section{Methods} \label{sec:mtds}
\subsection{Nearby and Distant component partitioning}\label{sec:mtds.partitions}
Figure \ref{fig:fieldloc_on_co} shows the distributions of Galactic polarization angles in each of the five fields under consideration. All fields show a clear limited range of angles, indicating a dominating regular field orientation along the LOS with small deviations. However, fields C11 and C15 appear to show a bimodality in their distributions.

A LOS can probe different interstellar structures located at different distances, which can produce the bimodality seen in fields C11 and C15. In this Section, we describe our procedure to disentangle possible different interstellar structures along the LOS probed by our five fields in and near the Coalsack.

In order to determine the number of discrete polarimetric components for each field, we fit different Gaussian distributions (single Gaussian curves, as well as the sum of two, three or four Gaussians) to the polarization angle histograms. The number of bins was determined using the Knuth rule \citep{knuth2006optimal, knuth2019optimal}, which is a data-based method that can be used to find the optimum number of histogram bins, based on Bayesian probability theory. The bin width of the histograms is thus different for each field. Using the Chi-squared value as a measure of goodness-of-fit, we are able to determine the number of components in each field, shown in Table \ref{tab:mastertable}. For fields C39, C40 and C54 a single Gaussian distribution provides a good fit. For fields C11 and C15, a model consisting of two Gaussians is more appropriate. We also perform an F-test \citep[][]{lupton1993statistics} and find that for fields C11, C15 and C40, it is warranted to include second Gaussian component to the model at the 99 per cent confidence level. The case of field C40 will be discussed in more detail in Section \ref{sec:disc.c40}.. Further analysis of the data, including partitioning, is done in the q, u, distance parameter space.

In the case of fields C11 and C15, stars belonging to either component are uniformly distributed across the field-of-view. It therefore seems unlikely that a small separate cloud or other dusty structure exists nearby in projection on the sky, but rather that one component is located at a different distance from the other. Indeed, this is confirmed by comparing the linear Stokes parameters q and u and the distances, see also Figures \ref{fig:qudist_3d} and \ref{fig:c11_QUavDist}. Therefore, partitioning of components based on q, u and the distance $d$ is justified. However, because of the unique characteristics of the three-dimensional q, u, d distributions for each field, we had to use different methods to partition the data into its discrete components, which will be explained in more detail in Sections \ref{sec:res.c11} and \ref{sec:res.c15}.

After partitioning, components are either assigned ``nearby" or ``distant" (N and D in Table \ref{tab:mastertable}, respectively). The nearby component consists of those stars at smaller distances where we can reasonably expect the Southern Coalsack to be the dominant polarizing feature. The light of these nearby stars is polarized by both the Southern Coalsack and, to a lesser extent, the LB wall. The distant stars are further away -- typically far behind the Coalsack cloud complex -- and show polarimetric signatures of at least two dominant polarizing structures along the LOS. 

In order to isolate the distant signal from the contributions of the nearby structure, characteristic values for the Stokes q and u parameters are determined for the nearby stars. These are the median values for q and u of all stars in the 80th distance percentile. Choosing the 80th percentile most distant stars of the nearby population ensures that we have a sizable representative sample. Using the most distant stars in the nearby population guarantees that we probe the entirety of the nearby ISM, including possible minor contributions from less dense dust behind the Coalsack. These proxy numbers are then subtracted from the distant population polarimetric measurements, resulting in foreground-corrected distant measurement (D$_{iso.}$ in table \ref{tab:mastertable}). We also correct for nearby contributions to the polarization angle dispersion, see Section \ref{sec:mtds.dcf} and Equation \ref{eq:poldisp_corr}. We can then determine the characteristics of the magnetic field for each component individually. Because the LB wall and Coalsack are likely attached \citep[see e.g.][and Figure \ref{fig:c11_QUavDist} below]{edenhofer20233ddustmap}, we are unable to disentangle contributions from the LB wall to the polarized signal of the nearby population of stars.

All average polarimetric values presented in Table \ref{tab:mastertable} are calculated from the weighted Stokes parameters, where the weights are set equal to the inverse of the associated errors. The distances and V-band extinctions are taken from \citet[][]{anders2022photo} and the averages of these values, taking their asymmetric errors into account, are calculated following the method described by \citet[][]{barlow2003asymmetric}, which was developed into a numerical method described in \citet[][]{laursen2019ADD_ASYM}.

\subsection{DCF parameters}\label{sec:mtds.dcf}
The DCF method \citep{davis1951strength, chandrasekhar1953magnetic} relates the magnetic field strength of the plane-of-sky magnetic field component (${B}_\perp$, hereafter B) to the polarization angle dispersion, gas velocity dispersion and gas mass density as follows:

\begin{equation}\label{eq:dcf}
    \textrm{B}_\perp = f \sqrt{4\pi\rho}\frac{\sigma_{v}}{\sigma_{\theta}},
\end{equation}

\noindent where $\rho$ is the local density of the gas (in $\textrm{g cm}^{-3}$) and $\sigma_{v}$ is the velocity dispersion of the gas (in $\textrm{cm s}^{-1}$). $\sigma_{\theta}$ refers to the dispersion in the polarization angles (in radians) and $f$ is a correction factor, meant to compensate for the fact that the DCF method generally overestimates magnetic field strengths. Numerical simulations of molecular clouds \citep[][]{ostiker2001molecularcloudmodels, heitsch2001magnetic} and protostellar cores \citep[][]{padoan2001modelspolarizeddustemission} show that $f=0.5$ gives reasonable estimates for the field strength..

We use CO maps to find DCF input parameters related to the specifics of the local ISM, $\sigma_{v}$ and $\rho$. CO maps in the general area of the Southern Coalsack are available as part of the \citet{dame2001completeCO} composite CO map. Specifically, the surveys of \citet{dame1987compositeco_dht01} (DHT01 in the composite map), \citet{nyman1989CO_dht35} (DHT35) and \citet{bronfman1989comolclouds_dht36} (DHT36) cover (part of) the region in which the observed fields can be found. The CO component associated with the Southern Coalsack has a velocity of $\sim-4\textrm{ km s}^{-1}$ \citep[][]{nyman1989CO_dht35, beuther2011coalsacknearfar}. For fields C15, C39 and C40, the typical Coalsack CO signature line at $-4\textrm{ km s}^{-1}$ is well visible above the noise in all three possible maps. Of these three options, the map by \citet{nyman1989CO_dht35} has the highest spatial and spectral resolution and is thus used. Although the $^{12}\textrm{CO}$ may be optically thick especially in high-density regions, \citet[][]{nyman1989CO_dht35} have also observed $^{13}\textrm{CO}$ in the Southern Coalsack and find that the line profiles of both species show great similarity \citep[see Figure 3 in][]{nyman1989CO_dht35} except in the most dense regions. Because all our fields are located away from the most dense regions, we are confident using $^{12}\textrm{CO}$ from \citet[][]{nyman1989CO_dht35} as a tracer.

Field C11 lies outside the area observed by \citet[][]{nyman1989CO_dht35}, see also Figure \ref{fig:fieldloc_on_co}. Field C54 lies on the edge of the area observed by \citet[][]{nyman1989CO_dht35} and is only partially covered. In this field, no feature at $-4\textrm{ km s}^{-1}$ is detected. Therefore, we cannot reliably determine the magnetic field strengths in the foreground of field C11 and the entire LOS of field C54.

For the fields where we do have reliable CO observations, a Gaussian fit to the spectrum at a velocity $-4\textrm{ km s}^{-1}$ will provide us with the velocity dispersion $\sigma_{v}$, which will be used in the DCF method.

The gas density $\rho$ is determined using the total CO intensity in the observed field. The intensity is converted to an $\textrm{H}_2$ column density using the \citet{bolatto2013conversion} conversion factor $X_{CO} = 2\times10^{20} \textrm{ cm}^{-2}\textrm{(K km s}^{-1}\textrm{)}^{-1} \pm 30\%$. Finally, we convert this value to a spatial number density by dividing the column density by the path length through the cloud. 

\begin{figure}[ht!]
\plotone{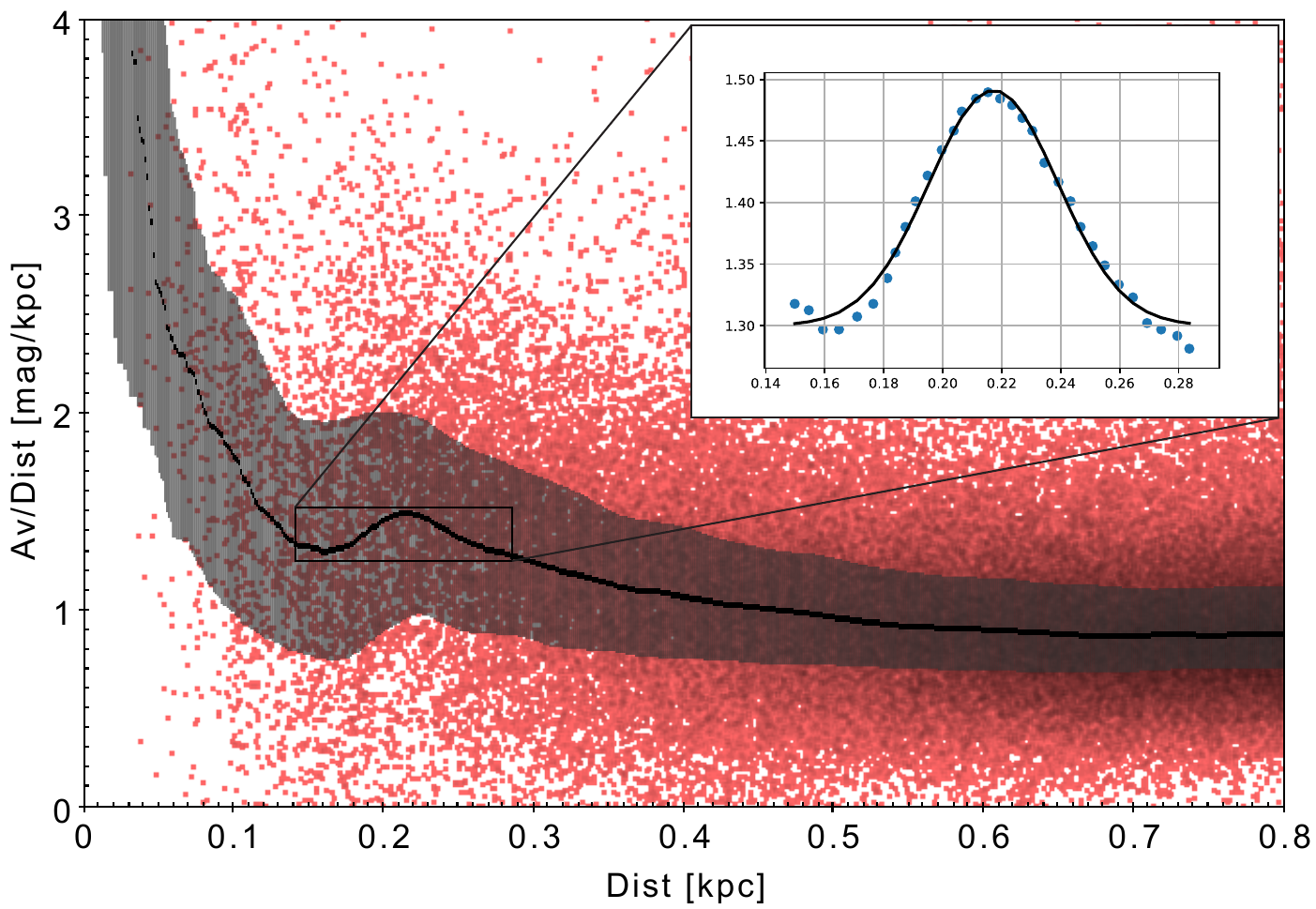}
\caption{Distance-normalized extinction ($A_{V}/dist$) versus distance for $\sim 2.5$ million stars in the general area of the Southern Coalsack. Data taken from \citet[][]{anders2022photo}. Red data points are the individual stars, the grey background indicates the 25\%-75\% quantile range and the black solid line indicates the median. The increase in $A_{V}/dist$ is caused by the presence of the Coalsack at a distance of $\sim185 \textrm{ pc}$. The inset shows a Gaussian fit to the $A_{V}/dist$ excess. We estimate a line-of-sight thickness of $25.7 \pm 0.6 \textrm{ pc}$.\label{fig:coalsack_thickness}}
\end{figure}

The path length is the thickness of the Southern Coalsack along the LOS. This path length was determined using the differential extinction ($A_{V}/d$) as a function of distance in GAIA data in the field. We downloaded all available data at the location of the Southern Coalsack in the \citet{anders2022photo} catalog (i.e. all available stars in the region $298\degr > l >308\degr$, $-4.5\degr > b > 2.5\degr$ for Galactic longitude and latitude, respectively) and plotted $A_{V}/d$ versus the distance $d$, see Figure \ref{fig:coalsack_thickness}. This shows an increase in differential extinction starting at $\sim 185 \textrm{ pc}$, which peaks a distance of $\sim 210 \textrm{ pc}$. The increase inDistances normalized extinction coincides with the near edge of the cloud (i.e. the distance to its near edge) and is in agreement with other measures of the distance to the Coalsack \citep[$d\sim180 \textrm{ pc}$, see e.g.][]{franco1989coalsack, seidensticker1989distancestructurecoalsack}. The far edge of the cloud is indicated by the peak in normalized extinction located at $\sim210 \textrm{ pc}$. Beyond this distance, the differential extinction decreases because there is little to no increase in $A_V$ per unit distance. It is thus expected that the cloud does not extend farther than the distance at which the peak of the excess $A_V/d$ is located. Fitting a Gaussian curve to the excess differential extinction, we can use the half-width at half maximum (HWHM) as an indicator of cloud thickness. We find that $l_{Coalsack} = HWHM = \sqrt{2ln2}\ \sigma = 25.7 \pm 0.6 \textrm{ pc}$. This is in agreement with other observations of 3D cloud structure, most recently by \citet[][]{dharmawardena3dmolclouds2023}, who find a thickness of $29 \textrm{ pc}$.

Finally, to convert the number density to a mass density, we assume the gas to be  90 \% molecular hydrogen ($\textrm{H}_{2}$) with a mass of $m_{H2} = 3.348\times10^{-24} \textrm{ g}$ and 10\% helium ($m_{He} = 6.646\times10^{-24} \textrm{ g}$) which will give us the local gas density $\rho$.

The final input parameter for the DCF method, the polarization angle dispersion $\sigma_{\theta}$, is corrected by the contribution to the observed dispersion originating from the observational errors using 
\begin{equation}\label{eq:polangdisp}
    \sigma_{\theta} = \sqrt{{\sigma_{\theta, G}}^2 - \langle{\sigma_{\theta, err}}\rangle^2},
\end{equation} 
\noindent where $\sigma_{\theta}$ is the polarization angle dispersion,  $\sigma_{\theta, G}$ is the standard deviation of the Gaussian fit to the polarization angle distribution (or some isolated component of the distribution) and $\langle\sigma_{\theta, err}\rangle$ is the mean error, following the method from \citet{pereyra2007polarimetry}. $\langle\sigma_{\theta, err}\rangle$ is typically small (using $\sigma_{\theta} = 0.5(\sigma_p/p)$ rad and an SNR of 5, the rms uncertainty is $\sigma_{\theta} < 5.8$ deg. The typical average error per field is around $\langle\sigma_{\theta, err}\rangle \sim 2 \degr$), which constitutes a small correction to the Gaussian fit.

For distant components, it is assumed that the ISM conditions reflect the canonical diffuse ISM values. Therefore, we adopt a value of $n(H_{2}) = 1 \textrm{ cm}^{-3}$ for the gas density \citep[see e.g.][]{ferriere2001interstellar}. For the velocity dispersion, we take the average value of the range $\sigma_{v} \sim 5-10 \textrm{ km s}^{-1}$ \citep[see e.g.][]{burkert2006turbulentISM, mckee2007theorystarformationreview, dobbs2011ismproperties} such that $\sigma_{v} = 7.5 \textrm{ km s}^{-1}$. Although it is possible, even likely, that the polarization of the distant stellar population is caused by a structure more dense than the diffuse ISM, such as dust clouds, filaments, or spiral arms, we currently do not know details about this secondary structure and therefore assume canonical diffuse ISM values. The validity of these assumptions will be discussed in more detail in Section \ref{sec:disc.bg}. 

In order to correct for contribution to the polarization angle dispersion of the far component by the nearby polarizing structure, we subtract in quadrature the nearby population's angle dispersion from the far population dispersion:

\begin{equation}\label{eq:poldisp_corr}
    \sigma_{\theta, D, corr} = \sqrt{\sigma_{\theta,D}^{2} - \sigma_{\theta,N}^{2}}.
\end{equation}

Equation \ref{eq:poldisp_corr} is a coarse approximation of correction for the contribution to the angle dispersion from a foreground polarizing structure. The resulting field strengths of the distant components will therefore be reported as order-of-magnitude estimates rounded to the nearest five $\mu$G because of the large uncertainties regarding the characteristics of the distant structures. 

The error on the plane-of-sky magnetic field strength (see equation \ref{eq:dcf}) will be calculated as:

\begin{equation}\label{eq:dcf_err}
    \frac{\delta B}{B} = \sqrt{ \left(\frac{1}{2}\frac{\delta \rho}{\rho}\right)^2 + \left(\frac{\delta \sigma_{v}}{\sigma_{v}}\right)^2 + \left(\frac{\delta \sigma_{\theta}}{\sigma_{\theta}}\right)^2 },
\end{equation}

\noindent derived from conventional error propagation. For the fields where CO maps are used to determine the gas density, 

\begin{equation}\label{eq:rho_err}
    \frac{\delta \rho}{\rho} = \sqrt{ \left(\frac{\delta W_{CO}}{W_{CO}}\right)^2 + \left(\frac{\delta X_{CO}}{X_{CO}}\right)^2 + \left(\frac{\delta l_{cloud}}{l_{cloud}}\right)^2 },
\end{equation}

\noindent where $W_{CO}$ is the total CO intensity, $X_{CO}$ is the CO-to-$\textrm{H}{_2}$ conversion factor \citep[][]{bolatto2013conversion} and $l_{cloud}$ is the thickness of the Coalsack. The input parameters for the DCF method and their associated errors (Eqs. \ref{eq:dcf} and \ref{eq:dcf_err}) can be found in Table \ref{tab:mastertable}.

Alternative methods for determining the magnetic field strength will be discussed in Sections \ref{sec:disc.othermethods} and \ref{sec:disc.sf_YA}.

\section{Results}\label{sec:res}
Because each field has different and unique properties, the analysis of the data differs from field to field. Therefore, we will discuss the results of each field separately.

\subsection{C11}\label{sec:res.c11}
\begin{figure}[ht!]
\centering
\includegraphics[width=0.5\textwidth]{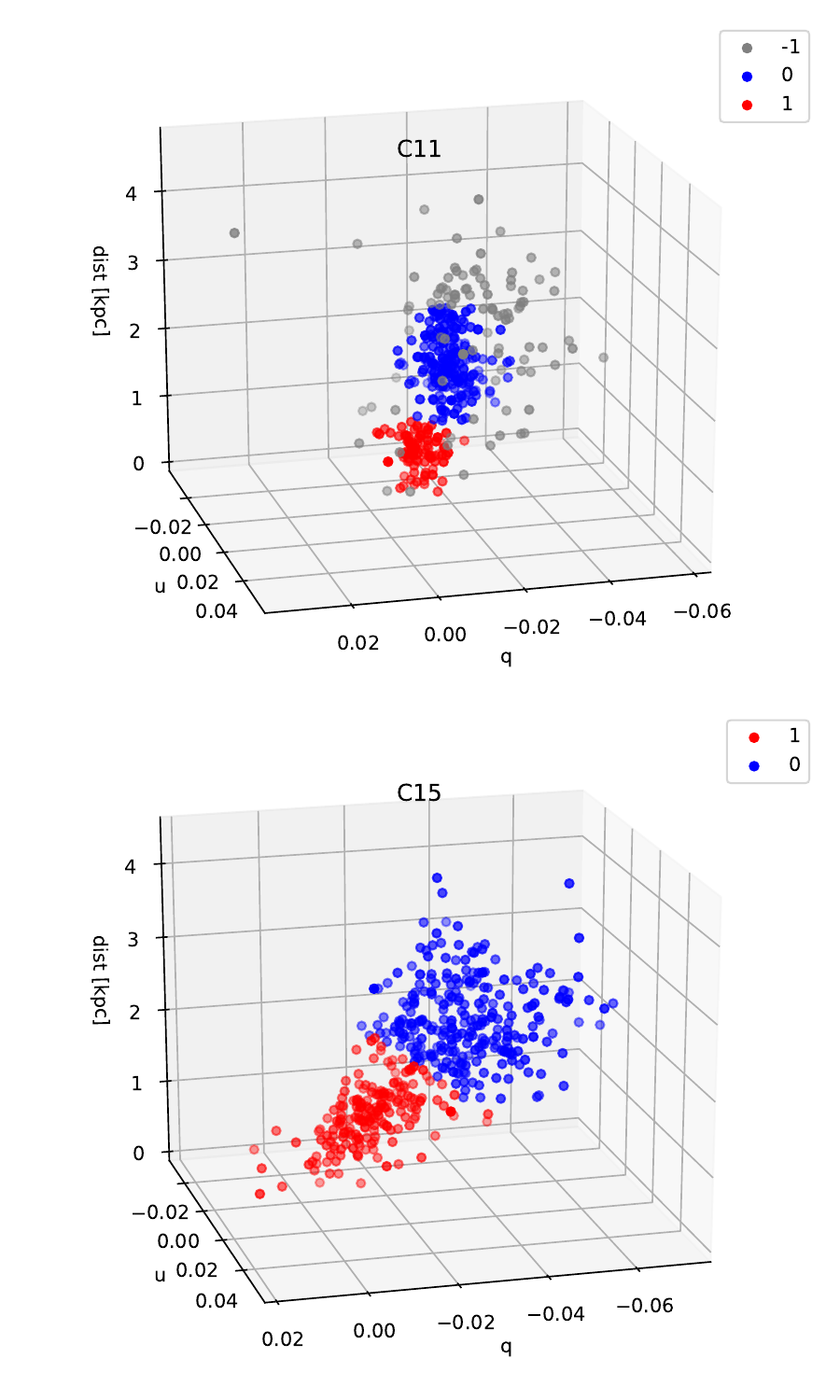}
\caption{Distributions of Stokes parameters q and u, and distance d in kpc. Top: field C11. Red and blue indicate the nearby and distant partitions, respectively, created using DBSCAN. Gray indicated data point that are not part of either component, see also Section \ref{sec:res.c11}. Bottom: field C15. Red and blue indicate the nearby and distant partitions, respectively created using K-means clustering; see also Section \ref{sec:res.c15}. \label{fig:qudist_3d}}
\end{figure}

Field C11 ($l=307.6\degr, b=1.3\degr$) contains 372 stars. This field is located outside the lowest CO contours at $2 \textrm{ km s}^{-1}$, see the circular marker in Figure \ref{fig:fieldloc_on_co}. Using multiple Gaussian fits, we find the distribution of polarization angles comprises two components, see Figure \ref{fig:fieldloc_on_co}, top left histogram, indicating at least two structures along the LOS. We first create two partitions to separate the data with polarization dominated by the nearby structure from that dominated by both structures, see also Section \ref{sec:mtds.partitions}. This particular field is very suited for partitioning using the \textsc{SciKit-learn} \citep{scikit-learn} DBSCAN algorithm \citep[][]{ester1996dbscan}{}{}. The algorithm finds clusters of overdensities in data separated by regions that are less densely populated. We applied this method to three parameters: the two linear polarization Stokes parameters q and u and the distance d, all normalized to the range 0-1. We set the DBSCAN parameters $\epsilon$ and $min_{samples}$ to $\epsilon=0.12$ and $min_{samples}=81$, respectively, based on silhouette score optimization. $\epsilon$ specifies the maximum distance between two data points beyond which they will not be considered part of the same cluster and $min_{samples}$ defines the minimum population of a dense region. The algorithm will use these two parameters to create the appropriate number of clusters in the data. In the case of field C11,  this creates two components of clustered data (the nearby and distant components) and a  component consisting of outlier stars that belong to neither cluster, see Figure \ref{fig:qudist_3d}, top, called ``noise" in the method \citep[][]{ester1996dbscan}{}{}. We emphasize that these data are high-quality, valid observations of stars, they just do not belong to the most dense clusters in the normalized q, u, d-space. We therefore do not apply further analysis of the magnetic field on the outlier points. 

Having created two distinct partitions in the data (the red and blue components in Figure \ref{fig:qudist_3d}, top), we use the method outlined in Section \ref{sec:mtds.partitions} to compute the polarization properties of the distant component. The polarimetric characteristics for each of the two components are presented in Table \ref{tab:mastertable}. The components are indicative of two distinct polarizing components along the LOS with very different magnetic field orientations. The nearby population, with an average distance of 900 pc, is dominated by a structure with a magnetic field orientation of $\theta_{N}=64\degr$, which we determine to be the Southern Coalsack. The average distance of the nearby partition extends far beyond the far edge of the Coalsack (i.e. $\sim 215$ pc, see also Section \ref{sec:mtds} and Figure \ref{fig:coalsack_thickness}), but the region immediately behind the Coalsack is mostly devoid of any dust \citep[][see also Figure \ref{fig:c11_QUavDist}]{mcclure2001hishells, nyman2008southerncoalsack}. The distant component, located at a much larger average distance of 2 kpc, shows a preferential magnetic field orientation of $\theta_{D}=84\degr$. After isolating the distant polarization from nearby contamination as outlined in Section \ref{sec:mtds.partitions}, the difference in orientation between the two components becomes even larger with $\theta_{D, iso.}=110\degr$. The turnover point between the two partition occurs at a distance of approximately $\sim 1.3 \textrm{ kpc}$, which is where we expect the secondary structure to begin. This dominant polarizing feature in the distant component will be discussed further in Section \ref{sec:disc.bg}.

Determination of the strength of the magnetic field requires certain information about the local ISM conditions. For field C11, no reliable estimates for the gas velocity dispersion and gas density are available in the \citet{nyman1989CO_dht35} CO map because the field lies outside the area observed. Therefore, we cannot determine the magnetic field strength of the nearby polarizing structure. 

The magnetic field strength of the distant structure, $B_{D, iso.}$, can be calculated using the characteristic diffuse ISM velocity dispersion and gas density values from Section \ref{sec:mtds.dcf}, i.e. $\sigma_{v} = 7.5 \textrm{ km s}^{-1}$ and $n(H_{2}) = 1 \textrm{ cm}^{-3}$, respectively. Using these values, the polarization angle dispersion as determined using Equation \ref{eq:polangdisp} and the DCF method (Equation \ref{eq:dcf}), we find a magnetic field with a strength of around $B_{D, iso.}\sim20 \ \mu \textrm{G}$.

\subsection{C15}\label{sec:res.c15}
Field C15, located at $l=304.7\degr$, $b=-0.2\degr$, is a moderately-densely populated field (461 stars) located near the edge of the Coalsack in CO. The distribution of polarization angles for field C15 shows two distinct components, see Figure \ref{fig:fieldloc_on_co}, bottom left histogram. As described in Section \ref{sec:mtds.partitions}, we want to create two partitions in the data. The three-dimensional distribution of q, u and distance show that these data are well-suited to be divided into two partitions using the \textsc{SciKit-learn} \citep{scikit-learn} K-means algorithm due to the overall low density and uniform distribution across the parameter space, and the similar size of the two potential components. We applied this algorithm to Stokes parameters q and u and the distance d. The K-means algorithm iteratively minimizes the Euclidean distance between each data point and an assumed centroid, updating the position of the centroid at each step until an optimal position is found. All parameters were normalized to the range 0-1 prior to partitioning to ensure an accurate result. The only input parameter for the K-means algorithm is the number $n$ of clusters that will be created. Having found that the best fit to the distribution of polarization angles has two Gaussian components, this is set to $n=2$.

Having created a nearby and distant component (the red and blue components in Figure \ref{fig:qudist_3d}, bottom, respectively), we applied the method described in Section \ref{sec:mtds} to isolate the distant stars from any polarizing contamination originating from the nearby cloud. The mean polarimetric characteristics for each of the components are presented in Table \ref{tab:mastertable}. The magnetic field in the nearby structure has a mean orientation of around $\theta_{N}=70\degr$, whereas the distant and isolated distant magnetic fields are oriented along $\theta_{D}=85\degr$ and $\theta_{D, iso.}=100\degr$, respectively. The boundary between the two polarizing regimes occurs at a distance of approximately $\sim 1.6 \textrm{ kpc}$. The nature of this secondary structure will be discussed in Section \ref{sec:disc.bg}.

We determine the input parameters for the DCF method (Equation \ref{eq:dcf}) using the methods described in Section \ref{sec:mtds.dcf}. The parameters used to determine the magnetic field strength can be found in Table \ref{tab:mastertable}. We find magnetic field strengths for the nearby and (isolated) distant structures of $B_{N}=12.5 \pm 2.1 \ \mu \textrm{G}$ and $B_{D, iso.}\sim10 \ \mu \textrm{G}$, respectively.

\subsection{C39}\label{sec:res.c39}
C39 is a field located near the Northern edge of the Coalsack at $l=303.1\degr, b=1.6\degr$, see Figure \ref{fig:fieldloc_on_co}, and comprises 309 stars. The fit to the polarization angle distribution for field C39 agrees best with a single-component Gaussian, see Figure \ref{fig:fieldloc_on_co}, bottom middle histogram. The mean polarimetric parameters of this single component are described in Table \ref{tab:mastertable}. We believe that this LOS is dominated by the Southern Coalsack nearby. The mean orientation of the magnetic field in this LOS is $\theta=83\degr$.

Next, we determined the parameters necessary to apply the DCF method following the methods described in Section \ref{sec:mtds.dcf}. The parameters are summarized in Table \ref{tab:mastertable}. We use those parameters and Equation \ref{eq:dcf} to determine the magnetic field strength and find $B=8.1 \pm 1.5 \ \mu \textrm{G}$ for field C39.

\subsection{C40}\label{sec:res.c40}
Field C40, located at $l=302.2\degr, b=-1.2\degr$, probes a more central region of the Southern Coalsack. This field contains 264 stars. The distribution of polarization angles for field C40 has been fit by a single Gaussian curve, see Figure \ref{fig:fieldloc_on_co}, bottom right histogram. We find that the chi-squared of the fit becomes slightly lower by adding a second Gaussian component centered around $\theta\sim96\degr$, but we are unable to attribute this component to an overdensity in either the q, u, distance parameter space, or the q, u, sky-coordinate parameter space. This will be explored further in the discussion in Section \ref{sec:disc.c40}. The single polarizing component along the LOS is oriented along an angle of $\theta=65\degr$.

The magnetic field strength along this LOS can be calculated using the polarization angle dispersion $\sigma_{\theta}$, gas velocity dispersion $\sigma_{v}$ and gas density $\rho$ as in Section \ref{sec:mtds.dcf} and are presented in Table \ref{tab:mastertable}. Using these values as input for Equation \ref{eq:dcf}, we find a magnetic field strength of $B=10.8 \pm 1.9 \ \mu \textrm{G}$.

\subsection{C54}\label{sec:res.c54}
Field C54 is located near the Northeastern edge of the Southern Coalsack ($l=305.2\degr, b=1.3\degr$, see also Figure \ref{fig:fieldloc_on_co}) and comprises 137 stars. The distribution of polarization angles is best characterized by a single Gaussian, see Figure \ref{fig:fieldloc_on_co}, center left histogram. The mean magnetic field orientation is $\theta=76\degr$. 

Determination of the magnetic field strength is made difficult by the fact that this field lies partially outside of the region observed by \citet[][]{nyman1989CO_dht35}. Only half of the field is covered in this CO map, see also Section \ref{sec:mtds.dcf}. Because no suffiecient CO maps are available with which to determine the local gas density and gas velocity dispersion, we are unable to calculate the magnetic field strength in this field.

\section{Discussion} \label{sec:disc}

\subsection{Possible secondary components}\label{sec:disc.c40}
The method of identification and isolation of different polarized stellar populations using clustering algorithms depends on being able to find those populations in q, u, distance parameter space. We describe some limitations of the method in two cases below.

A small change in angle may be visible in Figure \ref{fig:planck_pa}, field C39, between 1 and 2 kpc. This angle does not appear in the corresponding histogram (Figure \ref{fig:fieldloc_on_co}, bottom center), nor does it correspond to a cluster in q, u, distance-space (Figure \ref{fig:c11_QUavDist}). Therefore, we cannot identify this cluster in the data, and thus cannot quantify the effect is may or may not have on our results. However, if the angle change were real, we here interpret is as dispersion and the actual magnetic field strength would likely be higher.

As mentioned in Section \ref{sec:res.c40}, we find a possible secondary component in the polarization angle distribution for field C40. Adding a secondary component to the Gaussian fit lowers the $\chi^2$ from $\chi^2_{1 \textrm{ comp}} = 8.4$ to $\chi^2_{2 \textrm{ comps}} = 5.0$. The secondary Gaussian component has an average angle $\mu=95.8 \pm 1.9 \degr$ and a standard deviation of $\sigma=6.0\pm1.9\degr$. An F-test \citep[][]{lupton1993statistics} confirms that the addition of this second Gaussian component to the model is statistically warranted. However, we are unable to isolate this component for further analysis because we cannot correlate this component to a spatial parameter, neither the on-sky Galactic coordinates $l$, $b$ nor the distance $d$. Therefore, it is likely that the excess exists all throughout the observed field. Indeed, we find that the Planck polarization angle observed at the location of field C40 is $\theta_{Gal, Planck}=94.4\degr$, which is in agreement with the possible secondary Gaussian component. We therefore consider the possibility that a structure with a magnetic field orientation $\theta\sim95\degr$ exists farther along the LOS. This structure could either be a single polarizing dust structure (such as a cloud or arm) located at a distance beyond what is probed by optical starlight or a diffuse large-scale component that dominates the Planck signal (which probes the ISM along the entire LOS). Instead, the observed signal from starlight polarimetry probes only the local ISM and its magnetic field, which is dominated by the Southern Coalsack.

\subsection{Magnetic field orientations}
We now discuss the orientations of the various magnetic field components. As mentioned above, for the nearby population of stars, the polarization is induced by both the Southern Coalsack and the LB wall. However, because these two structures are attached \citep[][]{lallement20223dinterstellardust, vergely20223dextinction, edenhofer20233ddustmap}, it is not possible for us to disentangle the two contributions. Although field C11 does not show a CO feature that may be associated to the Coalsack, these dust models indicate that the LOS of field C11 has a Coalsack-like dust feature and thus may be considered part of the Coalsack. 

Firstly, the components that are dominated by the nearby structure (i.e.~C11, nearby component, C15, nearby component, and fields C39, C40 and C54) are generally oriented along an angle $\theta\sim60-80\degr$. This is in agreement with other measurements of the magnetic field orientation for the Southern Coalsack \citep[see][table 2 therein]{andersson2005highsampling}. 

In addition, the fact that the nearby component of field C11 (located outside the main CO emitting region) shows a similar orientation to the fields located within the CO contours and is in agreement with other findings \citep[see][Table 2]{andersson2005highsampling} provides evidence that the Southern Coalsack extends beyond the edges suggested by the CO emission \citep[][]{nyman1989CO_dht35}, see also Figure \ref{fig:fieldloc_on_co}. This is corroborated by the dust models of \citet{lallement20223dinterstellardust} and \citet{vergely20223dextinction}, who see Coalsack-like dust features in all five fields, see also the extinction profiles in Figure \ref{fig:c11_QUavDist}, as well as \citet{andersson2004hotenvelope}, who observe a large hot envelope surrounding the Coalsack.

Secondly, the distant structure-dominated components (i.e. C11, distant component, and C15, distant component) are oriented along angles that are higher compared to the foreground: $\theta\sim100-120\degr$. The polarization angle of the distant component could be caused by the large-scale, regular GMF component in the Milky Way. A toy model GMF in the Galactic plane, following the spiral arms, would show a polarization angle $\theta = 90 \degr$. However, more complex models of the GMF show out-of-plane components \citep[][]{jansson2012model, rodrigues2019evolution,shukurov2019modelling}, and so do infrared polarimetric observations in the plane in the first quadrant \citep[][]{clemens2020gpipsdr4}. Indeed, the observations presented above show a departure from $90\degr$ toward higher polarization angles. This distant component will be discussed in more detail in Section \ref{sec:disc.bg}.

\subsection{Magnetic field strengths}
For two fields (C11 nearby and C54), we are unable to accurately determine the magnetic field strength because these fields lie (partially) outside of the observed area by \citet{nyman1989CO_dht35}. To get an order-of-magnitude idea of the field strength in those regions, we use the average gas density and gas velocity dispersion from the fields where we do have a CO detection (C15 nearby, C39 and C40). For the foreground of field C11, we find then a magnetic field strength for the nearby component of $B_{N}=27.1 \pm 5.2\textrm{ }\mu \textrm{G}$. For field C54, we extrapolate the available data to arrive at an upper limit for the gas density $\rho < 0.9\pm0.3 \times 10^{-22} \textrm{ g }\textrm{cm}^{-3}$. Because no Coalsack signature is detected, we use the average value from fields C15 (nearby), C39 and C40 for the gas velocity dispersion to arrive at a magnetic field strength of $B<6.1 \pm 1.2 \ \mu \textrm{G}$. These values are speculative but in rough agreement with the findings in the other fields.

Firstly, we find that components dominated by the Southern Coalsack (i.e. C11 (nearby component), C15 (nearby component), C39, C40 and C54) have plane-of-sky magnetic fields with a strength ranging between $B\sim 8 \ \mu \textrm{G}$ and $B\sim 25 \ \mu \textrm{G}$, with an average nearby field strength of $B\sim 13 \ \mu \textrm{G}$. As the magnetic field component parallel to the LOS is unknown, this constitutes a lower limit to the total magnetic field strength. Even though Zeeman measurements probe the parallel component of the magnetic field, large statistical samples of Zeeman measurements can give an idea of typical field strengths in clouds. \citet[][]{myers1995magneticdiffuse} find parallel magnetic field strengths in diffuse clouds from 4 to~19 $\mu$G, while \citet[][]{crutcher1999magfieldmolclouds} discusses field strengths in 27 molecular clouds and reports values from a few $\mu$G to 3.1 mG. The observed perpendicular field strength in the Southern Coalsack is consistent with these values.

Furthermore, we find that our magnetic field strengths $B\sim10 \ \mu \textrm{G}$ are broadly in agreement with other measurements of the strength of the perpendicular component of the magnetic field in the Southern Coalsack-- differences between our findings and those from literature can largely be attributed to differences in methodical choices. \citet{lada2004dustyring}, who base their calculations on observations by \citet{jones1984innercore}, find a magnetic field strength of $B=23 \ \mu \textrm{G}$. A re-analysis of \citet{lada2004dustyring} by \citet{bhat2011bfielducl}, using the same observations from \citet{jones1984innercore}, finds a magnetic field strength of $B\sim32 \ \mu \textrm{G}$. \citet{andersson2005highsampling} calculate a magnetic field strength of $B=93 \ \mu \textrm{G}$ and have corrected the magnetic field strength of the Coalsack for foreground contributions to the polarization angle dispersion using polarimetric radiative transfer. Without this correction, the magnetic field strength is found to be $B=53 \ \mu \textrm{G}$ These results are in rough agreement with our findings, taking into consideration that we have used a correction factor f=0.5 in our application of the DCF method.

Secondly, we find that distant components, after isolation from nearby contributions, have magnetic field strengths of around $B\sim10-20 \ \mu\textrm{G}$. These values are found using canonical values for the diffuse ISM for the gas density $\rho$ and velocity dispersion $\sigma_{v}$ and will be discussed in more detail in Section \ref{sec:disc.bg}.

We note that \citet[][]{andersson2005highsampling} see signatures of the LB wall in their analysis of the polarimetric data which has a significant effect on their determination of the magnetic field strength of the Coalsack. However, they do not see LB Wall contributions in all of their fields. It is likely that the contribution of the LB wall to the polarized signal is not uniform across the sky. Due to varying dust contents and changing magnetic field geometries  \citep[][]{bailey2016lbdibs, medan2019lbw, pelgrims2020modelinglbmag}, the LB wall contribution could be significant in some areas and negligible in others. Because we cannot distinguish any contribution from the LB wall, we are unable to accurately assess the contribution of the LB wall to the polarimetry of the observed stars. It is also possible that the edge of the Southern Coalsack coincides with the LB wall. This is strengthened by the fact that in extinction, we do not see separate components for the LB wall and the Coalsack, see e.g. Figure \ref{fig:c11_QUavDist}, bottom line plots. This means that we are unable to disentangle contributions from the LB wall and the Coalsack.

We emphasize that IPS-GI fields are located away from the most dense regions of the Coalsack (see Figure \ref{fig:fieldloc_on_co}). As such, the magnetic field orientations and strengths may be accurate for those specific locations, but might not be representative of the entire Southern Coalsack.

We correct polarization angle dispersion of the distant populations by subtracting in quadrature the nearby angle dispersion (Equation \ref{eq:poldisp_corr}). This approximation is necessary because we lack precise details about the local ISM conditions of the distant polarizing structure. If those details were known, it would be valuable to apply a more detailed method, such as a full polarized radiative transfer analysis \citep[see e.g.][]{andersson2005highsampling}

Finally, we note that the errors in the magnetic field strength are dominated by the uncertainties in the gas density, especially due to the error in the CO-to-$\textrm{H}_{2}$ conversion factor \citep[][]{bolatto2013conversion}. The error in density makes up between 80-90\% of the total error in magnetic field strength. However, because of the location of the fields (in the Galactic plane, in the Southern sky) very few other tracers  of a similar scale and resolution to DHT35 are available. There are some observations of stars near the Coalsack. \citet{andersson2002ultrahighCH} observe CH in three stars close to the Southern Coalsack and \citet{corradi2004localISMcoalsack} observe a further four stars in Na I. These observations show velocity dispersions a bit larger than our findings using CO, but they do not cover the same area that is observed in CO and specifically not the regions of the Coalsack where our observed fields are located. 

Summarizing, various estimates for magnetic field strengths in the Coalsack are quoted in the literature. However, a detailed look at the distinct methods of analysis shows that there are a number of valid reasons why differing field strengths are derived, such as the inability to separate the LB wall contribution, analyses that only look at certain parts of the Coalsack and inclusion of correction factors in the DCF method or not.

\subsection{Comparison with the Heiles catalog}
The agglomeration of polarimetric catalogs by \citet{heiles20009286} also contains stars located in or close to the Southern Coalsack. We have identified 84 stars \citep[taken from][see also \citet{heiles20009286} for more details and a quantitative comparison between these data sources]{vansmith1956interstellarpol, mathewson1970polobs1800stars, klare19771660southOBstarpol, reiz1998ubvpolselectedareas} that are located within the lowest boundaries in the CO maps of \citet{nyman1989CO_dht35}. Using these polarimetric observations, we find average polarimetric values of $p = 1.2 \pm 0.07\%$ and $\theta_{gal} = 86.3 \pm 1.6 \degr$. We use the mean values for $\rho$ and $\sigma_{v}$ to compute a magnetic field strength of $B = 12.2 \pm 2.3 \ \mu\textrm{G}$ using the DCF method and $B = 7.0 \pm 1.2 \ \mu\textrm{G}$ using the ST method. The magnetic field orientation and strength are in agreement with field strengths determined based on IPS data (Table \ref{tab:mastertable}). We have identified one star in common between the two catalogs: HD 113754, located in IPS field C15 at a position $l=304.71\degr, b=-0.17\degr$. We consider this star to be a background star at a distance $d=1.9 \textrm{ kpc}$. The polarimetric parameters taken from both \citet[][]{heiles20009286} and the IPS catalog are in rough agreement: the degree of polarization is $p=4.1\pm 0.2\%$ and $p=5.2\pm 0.05\%$ in the former and latter catalogs, respectively. The polarization angle is $\theta_{Gal}=76\pm 1.1\degr$ and $\theta_{Gal}=76.4\pm 0.3\degr$ for \citet[][]{heiles20009286} and IPS, respectively.

\subsection{Comparison with Planck observations}
\begin{figure}[ht!]
\centering
\includegraphics[width=0.5\textwidth]{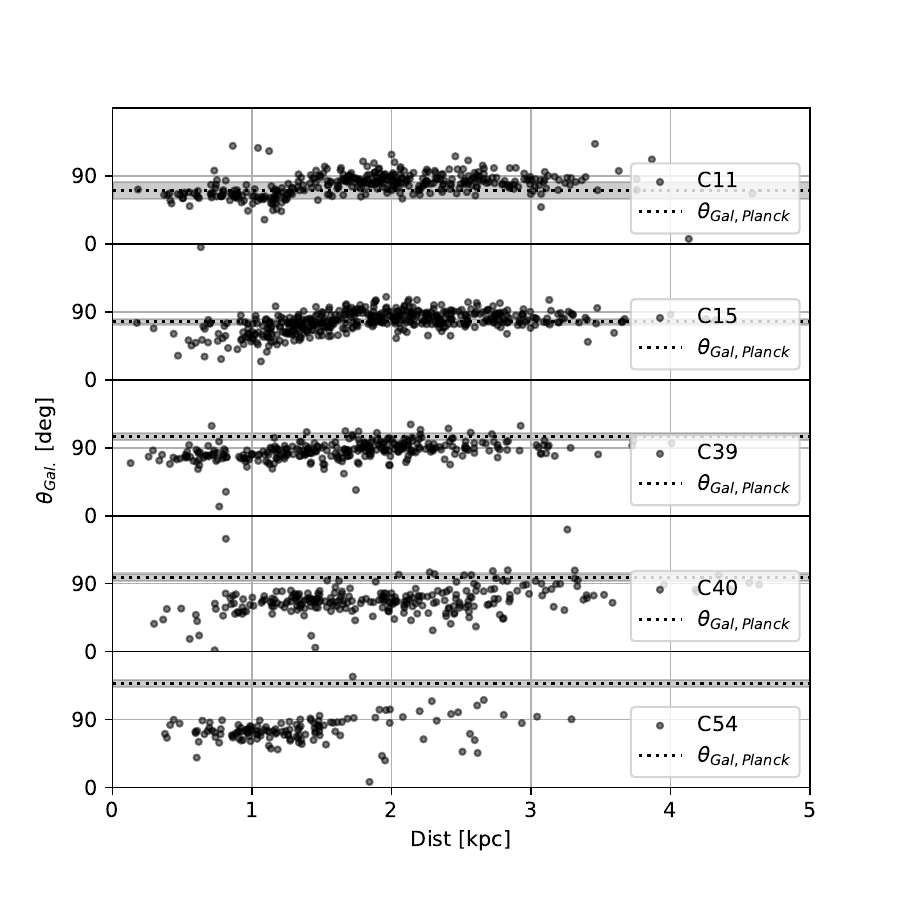}
\caption{Per field, for each star its polarization angle and distance. Distances taken from \citet[][]{anders2022photo}. The dotted lines indicate the average magnetic field orientation at the location of the center of the field taken from the Planck 353 GHz HFI maps. The shaded area shows the uncertainty of the Planck polarization angle. \label{fig:planck_pa}}
\end{figure}

Observations of polarized dust emission by Planck provide another auxiliary measurement to compare our findings to. We use the thermal dust maps at variable resolution to determine a single orientation for the magnetic field as measured by Planck. In Figure \ref{fig:planck_pa} we plot the polarization angle from the IPS-GI observations with their distance. The dotted lines indicate the magnetic field orientation as measured by Planck. Figure \ref{fig:planck_pa} shows that Planck sometimes measures a field parallel to the nearby structure (i.e. the Southern Coalsack) as in field C11, it sometimes is parallel to the distant structure (as in field C15) and it sometimes finds a different orientation altogether (as in fields C39, C40 and C54).

Despite this, we do find some consistencies. Firstly, in fields where IPS-GI find more than one polarimetric component (fields C11 and C15), Planck finds a magnetic field orientation $\theta_{Gal} < 90\degr$, indicating that the nearby component is a significant contribution to the Planck polarization. Conversely, in fields with only one component, Planck generally finds orientations $\theta_{Gal} > 90\degr$. Planck appears to pick up on a dominant background that lies beyond the distances probed by starlight polarization in certain fields, see also Section \ref{sec:disc.c40}.

For fields C15, C39 and possibly C40, we see that with increasing distance, there is a closer agreement between the optical and submillimeter polarization angles. This is likely a sign that in those fields, the most distant stars trace the same dust as Planck does and that there is little dust contributing to the polarized submillimeter signal beyond that distance.

\subsection{Structure functions of polarization angles} \label{sec:disc.sf_YA}
The DCF method derives the field strength from the dispersion of the polarization angles, the displacement of which is assumed to be caused by the propagation of incompressible transverse magnetohydrodynamic (MHD) waves known as Alfv\'en waves. The DCF method assumes that the polarization angle dispersion measured across the field of view is representative of the total turbulent angle dispersion. Therefore, if the largest turbulent scale is larger than the field of view, the measured angle dispersion is a lower limit and the derived magnetic field is an upper limit to the true value. On the other hand, if the maximum turbulent scale is smaller than the resolution, the angle dispersion is solely caused by large-scale variations in the regular magnetic field component. In such a case, the magnetic field strength can be obtained using the method described by  \citet[][]{hildebrand2009dispersion}, who propose using a more accurate way to calculate the plane-of-sky turbulent dispersion of local structured fields (e.g. in clouds) via the structure function of the polarization angles, the Angular Dispersion Function (ADF):

\begin{equation}
\label{eq:struc_func_PAdiff}
    \langle\Delta\theta^2(\ell)\rangle^{1/2} = \left( \frac{1}{N(\ell)}\sum^{N(\ell)}_{i=1} [\mathbf{\theta(x)-\theta(x+\boldsymbol\ell)}]^2\right)^{1/2} ,
\end{equation}

\noindent where $\theta(\mathbf{x})$ is the polarization angle at the position $\mathbf{x}$ and $\ell$ is the angular separation between two sources in the sky. This method does not need to assume a model of the magnetic field and provides an independent measurement of the turbulent to large-scale magnetic field strength ratio, $\langle B_t^2\rangle^{1/2}/B$ \citep[see][for more details]{hildebrand2009dispersion}. 

\begin{figure*}[ht!]
\plotone{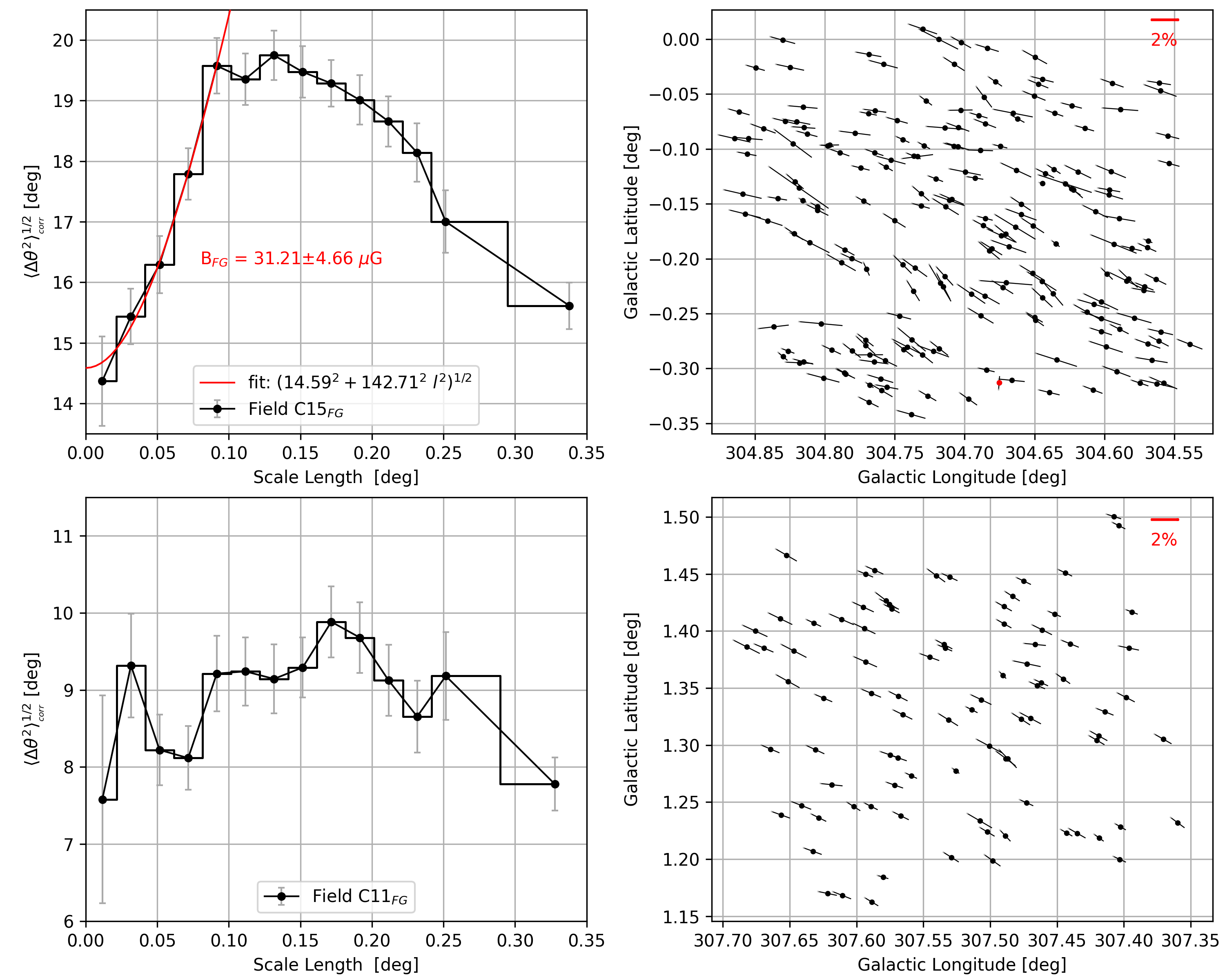}
\caption{Top left: the square root of the structure function of the polarization angles (the ADF, see Eq. \ref{eq:struc_func_PAdiff}) of the nearby population of field C15. The red line indicates the best fit to the first five bins. Top right: polarization vector plot of the nearby stars of field C15. The red marked star indicates an observation removed from further analysis. Bottom left: the ADF (Eq. \ref{eq:struc_func_PAdiff} of the polarization angles of the nearby population of field C11. Bottom right: polarization vector plot of the stars in the nearby population of field C11. \label{fig:strucfuncs}}
\end{figure*}

We use Equation \ref{eq:struc_func_PAdiff} to find the ADF in the nearby components of the Coalsack's IPS-GI regions, i.e. C11 and C15 (see Figure \ref{fig:strucfuncs}). For these two nearby components, we can apply the structure function method because we can approximate that part of the LOS as a single polarizing dust sheet with a population of stars, all with similar polarimetric properties, located immediately behind the dusty screen. This simplification is necessary because we use the plane-of-sky position to determine the scale length in the structure function, disregarding the distance between the stars. Because all nearby stars are polarized by the Southern Coalsack only, and there are no other dusty structures that may add to the polarization of the starlight, we can safely make this simplification. This also explains why we cannot readily apply this method to the other LOSs or the distant components of fields C11 and C15-- we cannot simplify those LOS to a single polarizing dust screen with a single stellar population with similar polarimetric properties, because there are multiple polarizing structures along the LOS.

Figure \ref{fig:strucfuncs} shows the ADF for the nearby populations of fields C11 and C15. Starting with field C15, the ADF increases with the observed scale up to $\ell\sim0.1\degr$ and then flattens. In field C11, on the other hand, the ADF is basically constant throughout the observed scales. Due to the angular size of the field-of-view, which is 0.3\degr \space by 0.3\degr, we only trust scales up to $\sim0.2\degr$, above which the statistics may be inaccurate. This is also true for the first scale bin.

There are different ways we can interpret the ADFs of Figure \ref{fig:strucfuncs}. Firstly, if the turbulent scale $\delta$ is larger than the smallest scale we can probe, i.e. $\delta>\ell_{min}$ with $\ell_{min}\sim0.01\degr$, and smaller than the large-scale variations of the magnetic field, $d$, the ADF will increase linearly (in log-log space) with the smallest scales until it reaches the outer scale of the turbulence. In this case, we would be observing a turbulent scale of 0.35 pc at the assumed cloud distance ($\sim 200$ pc) in field C15, but we can not calculate the magnetic field strength.

Secondly, if the turbulent scale is smaller than the smallest scale we can probe, i.e. $\delta<\ell_{min}$ such that $\delta < \ell \ll d$, then $\ell_{min}$ is the turbulence scale upper limit and the ADF can be fitted at the smallest scales as described in \cite{hildebrand2009dispersion}:

\begin{equation}
\label{eq:struc_func_fit}
    \langle\Delta\theta^2(\ell)\rangle_{tot} = b^2 + m^2\ell^2 + \sigma^2_M(\ell)  ,
\end{equation}

where $\langle\Delta\theta^2(\ell)\rangle_{tot}$ is the total function calculated from the observations, $\sigma^2_M(\ell)$ are the measurement uncertainties that must be subtracted to obtain the values of Equation \ref{eq:struc_func_PAdiff}, $b$ is the turbulent contribution, and $m\ell$ is the contribution of the large-scale fluctuations. According to \cite{hildebrand2009dispersion}, $b^2$ is proportional to $\sigma_{\theta}^2$, which is also assumed to be approximately $\delta B/B$, the variation in the magnetic field about its large-scale component. Therefore, we can calculate the the regular  magnetic field strength as

\begin{equation}
\label{eq:Hild_B0}
    B = \sqrt{8\pi\rho}\frac{\sigma_v}{b} ~~~~ \mathrm{for} ~~~~ B_t \ll B,
\end{equation}

\noindent where $B_t$ is the turbulent magnetic field strength and the turbulent velocity dispersion, $\sigma_v$, and density, $\rho$, of the cloud must be very well known. A correction factor f as in Equation \ref{eq:dcf} is not required. Then, if this is the case for field C15, the structure function fit gives $b = 14.59\degr\pm0.13\degr$ (the red line in Figure~\ref{fig:strucfuncs}, top row), and the magnetic field strength would be $31.2\pm4.7 \ \mu$G considering the parameters of Table 1.

Finally, we find that the structure function of C11 (Figure~\ref{fig:strucfuncs}, bottom left) is constant (i.e. fitting the structure function with a parabola as in Field C15 versus a straight horizontal line results in similar $\chi^2$-values). This suggests that the large-scale variations are not significant and the turbulent outer scale is below the minimum scale we can probe, $\ell_{min}$. This means that the magnetic field is significantly regular as it can also be observed in the polarization vector map (Figure \ref{fig:strucfuncs}, bottom right) and the polarization angle histogram for this field, Figure \ref{fig:fieldloc_on_co}, top left. 

\subsection{Distant structure}\label{sec:disc.bg}
\begin{figure*}[ht!]
\plotone{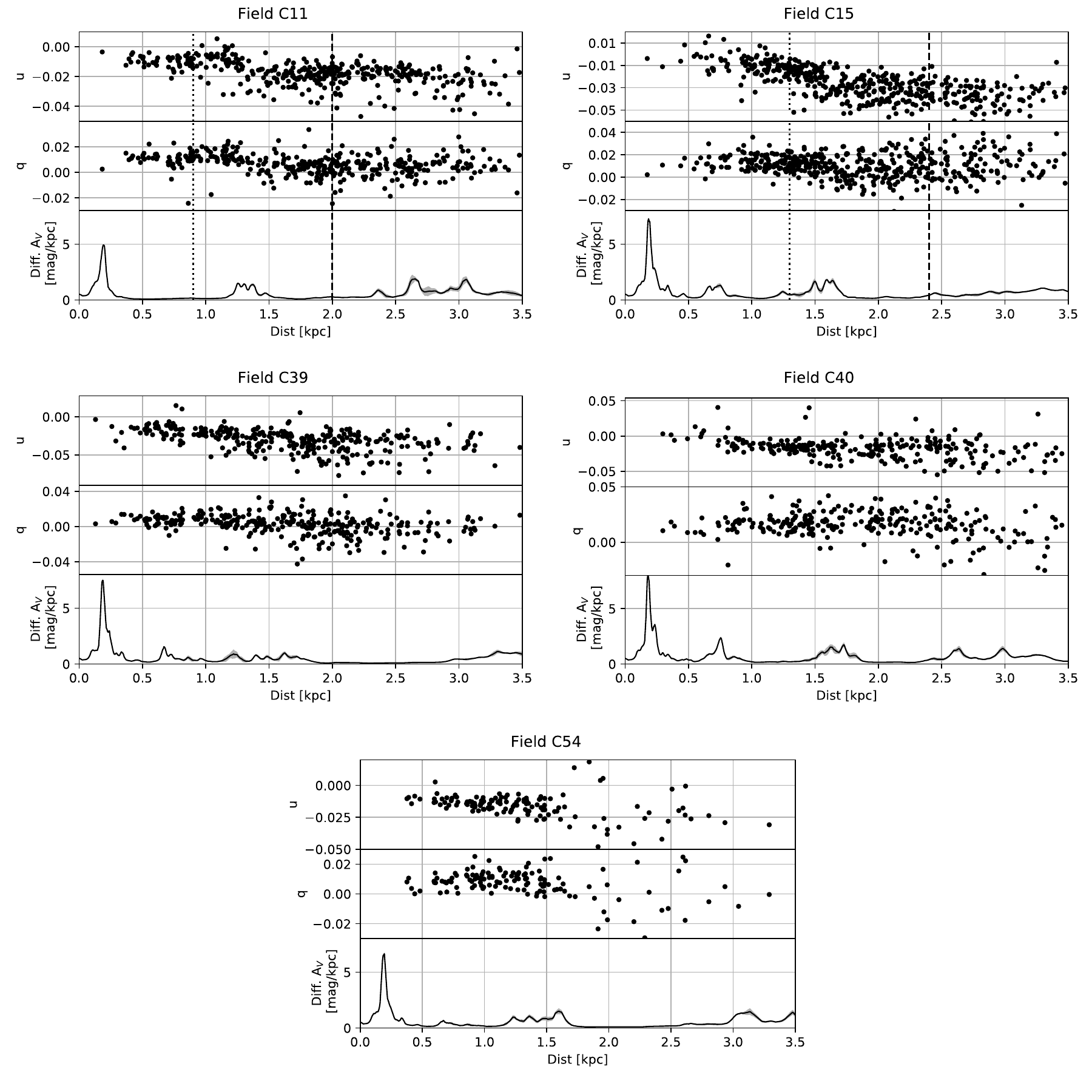}
\caption{Scatter plots of Stokes parameters q (top panel) and u (middle panel) versus distance for all fields. The bottom profiles show the differential extinction (or extinction density) in mag/kpc taken from \citet[][]{lallement20223dinterstellardust, vergely20223dextinction}. The dotted and dashed vertical lines in the scatter plots of fields C11 and C15 indicate the average distance of the nearby and distant populations, respectively. The turnover point at which the polarized signal changes occurs at a distance of approximately 1.3 kpc. \label{fig:c11_QUavDist}}
\end{figure*}

The observations of fields C11 and C15 (see Sections \ref{sec:res.c11} and \ref{sec:res.c15}, respectively) indicate lines-of-sight with at least two polarizing structures. This is most clearly evident when we plot the Stokes parameters q and u as a function of distance as in Figure \ref{fig:c11_QUavDist}, top left and right. Figure \ref{fig:c11_QUavDist}, top left and right shows a somewhat constant value for q and u up to a certain distance, beyond which the Stokes parameters assume a different value.  The distance at which this happens is around $d\sim1.3 \textrm{ kpc}$ for field C11, and $d\sim1.5 \textrm{ kpc}$ for C15. We interpret this as follows: starting from Earth, the LOS first penetrates the Southern Coalsack. This is a dense, dusty structure which causes a polarized signal. This affects the stars within the Coalsack, but also those behind the cloud complex. The region immediately behind the Coalsack appears to be mostly devoid of (dense) dust \citep[][]{seidensticker1989distancestructurecoalsack}, see also Figure \ref{fig:c11_QUavDist}. This means that stars located in this low-density region will exhibit a polarized signal dominated by the Southern Coalsack nearby. This is what is represented in Figure \ref{fig:c11_QUavDist} at distances $d<1.5\textrm{ kpc}$.

Around $d\sim1.5\textrm{ kpc}$, a change in polarimetry is observed. This must be caused by a second polarizing component along the LOS, indicating another dusty, dense structure capable of dominating the polarimetry of the stars. There are various possibilities of what precisely constitutes this structure.

Firstly, the CO observations of \citet{beuther2011coalsacknearfar} show not only the well-known Coalsack signal at $-4 \textrm{ km s}^{-1}$, but also a second component at a velocity of $-30 \textrm{ km s}^{-1}$. \citet{beuther2011coalsacknearfar} put this component at a distance of $d=3.1 \textrm{ kpc}$, which is too great a distance to account for the changes in polarimetry seen in Figure \ref{fig:c11_QUavDist}. Therefore, we dismiss this possibility. 

Secondly, there is evidence of a very large HI shell (GSH 304-00-12) located at a distance of about $d\sim1.2 \textrm{ kpc}$ behind the Southern Coalsack \citep[][]{mcclure2001hishells}. This structure is located at the exact distance at which we see a change in polarimetry in fields C11 and C15.  However, this HI shell is very large (approximately 20\degr \space by 30\degr) which means we should be able to see this shell in the distant components of all the IPS fields, assuming that the dust in the HI shell is distributed uniformly enough to be detectable throughout the shell. However, we do not see similar changes in the polarimetry of fields C39, C40 and C54. 

The final possibility is that the starlight from the distant stars is polarized by dust in the Carina spiral arm \citep[see also][]{seidensticker1989distancestructurecoalsack} as suggested by \citet{andersson2005highsampling}, who see the spiral arm in the polarimetric signature of one of their components (their component 2). Their magnetic field orientation of this component, described as the Carina arm, is $\theta = 88.5\degr$ on average, which is a bit lower than our measurements for the orientation (i.e. $\theta \sim100-120\degr$). However, they, as well as e.g. \citet{seidensticker1989distancestructurecoalsack} estimate a distance to the Carina arm of $d=1.3 \textrm{ kpc}$ which does coincide with our estimates of the distance of the second polarizing structure. However, because of the large size of the spiral arm, we would expect to see signatures of the Carina arm in the polarimetry of most, if not all, observed fields-of-view. Therefore, this distant component is believed to be a structure at roughly the Carina arm's distance, likely inside the arm.

Our detections of the distant structure rely on a change in polarization observed along the LOS. If the secondary structure were to have similar magnetic field properties as the nearby structure, the method described above would not be sensitive to the distant polarizing screen. In that case, a more detailed analysis of dust models would be warranted.

We note that the two fields that show a strong secondary polarizing component along the LOS, i.e. fields C11 and C15, are both located toward the Northeast of the Southern Coalsack. Therefore, it is possible that the second structure only exists in that region, that its dust content is limited to that region or that outside of that region its magnetic field properties are more similar to that of the Coalsack in front. However, we note that these fields are located near the outskirts of the Coalsack. This means that polarimetric signal of these specific lines-of-sight may be less dominated by the Coalsack. For example, the dust content along the edges of the Coalsack may be less than in more dense regions toward the center. This would allow a secondary structure to be more visible in the observations. More observations are necessary to fully confirm the nature of this secondary structure.

\subsection{Other methods of field strength determination}\label{sec:disc.othermethods}
In the analysis presented above, we have chosen to use the DCF method \citep[][]{davis1951strength, chandrasekhar1953magnetic}, see also Equation \ref{eq:dcf}, to determine the strength of the magnetic field components. Although this method is very well-established and many adaptations and corrections to the method have been suggested, it is known that the DCF method overestimates the magnetic field strength if a correction factor is not applied. Therefore, it is desirable to compare different methods of field strength determination.

An alternative to the DCF method based was presented by \citet{skalidis2021strengthestimation}, hereafter ST, who have relaxed the assumption of incompressibility necessary for the DCF method and instead derived:

\begin{equation}
    B = \sqrt{2\pi\rho} \frac{\sigma_{v}}{\sqrt{\sigma_{\theta}}}
\end{equation}

\noindent where $\rho$ is the gas density, $\sigma_{v}$ is the velocity dispersion and $\sigma_{\theta}$ is the polarization angle dispersion. We have also used the ST method to calculate the strengths of the various magnetic field components. The results can be found in Table \ref{tab:mastertable}. The ST method finds lower magnetic field strengths than the DCF method. The differences are largest for the strongest magnetic fields (i.e. C15 (nearby) and C40). Overall, we find that using the ST method, magnetic fields are found to be around 50\% weaker compared to the DCF method.

Another method that can be used to determine magnetic field strength is the Differential Measure Analysis (DMA), see \citet[][]{lazarian2022magnetic}. This method uses gradients (in the form of structure functions) of the gas velocity and polarization angle dispersion and is based on MHD turbulence theory. \citet[][]{lazarian2022magnetic} estimate the value of the correction factor f for different turbulent regimes. However, the required properties, such as Mach numbers and plasma $\beta$, are not well known in the Coalsack. However, if we assume low-$\beta$ conditions in the Coalsack (as in their Table 1) and sub-alfvenic turbulence \citep{li2013magneticfieldsfilamentaryclouds}, Figure 13 in \citet{lazarian2022magnetic} shows that the factor f can range from 0.2 to 0.9 depending on Alfvenic mach number and line-of-sight angle of the local magnetic field. In addition, we do not have detailed enough information to construct a structure function of gas velocities across the fields-of-view. As these quantities are either unknown or very uncertain, we do not have sufficient information to further constrain the magnetic field in the Coalsack with this method.

\section{Summary and Conclusions} \label{sec:con}
We have presented optical V-band observations of starlight polarization from IPS-GI in five discrete fields-of-view in or close to the Southern Coalsack. Combining these observations with accurate distance measurements, we are able, in two of our fields, to disentangle starlight polarization dominated by the Coalsack and a polarimetric signature of a dusty structure behind the Coalsack, at a distance of roughly $\sim 1.3-1.5 \textrm{ kpc}$. Using the K-means clustering and DBSCAN algorithms and by subtracting the nearby polarimetric contribution from the distant polarization, we are able to study both structures independently. For the Coalsack-dominated nearby population, we find magnetic field orientations in the range $\theta_{N}=65-85\degr$. For the polarizing distant structure, we find a plane-of-sky magnetic field orientation of $\theta_{D}=100-110\degr$. Furthermore, using CO observations of \citet[][]{nyman1989CO_dht35}, we can use the DCF method to find an estimate of the strength of the plane-of-sky component of the magnetic fields of the nearby and distant structures. The strengths we find are in the range $B=8-12 \ \mu\textrm{G}$ and $B\sim10 \ \mu\textrm{G}$ for the nearby and distant structures, respectively. The magnetic field strength of the distant structure is an order-of-magnitude estimation due to uncertainties in its local gas density and gas velocity dispersion. Further observations may help uncover its precise nature and aid in improving the field strength estimation.

\begin{acknowledgments}
The authors are very grateful to the reviewers, whose helpful suggestions and insights helped us improve the manuscript.
MJFV acknowledges Wouter Veltkamp for his contributions to Figure \ref{fig:fieldloc_on_co}. 

Over the years, IPS data has been gathered by a number of dedicated observers, to whom the authors are very grateful: Flaviane C. F. Benedito, Alex Carciofi, Cassia Fernandez, Tibério Ferrari, Livia S. C. A. Ferreira, Viviana S. Gabriel, Aiara Lobo-Gomes, Luciana de Matos, Rocio Melgarejo, Antonio Pereyra, Nadili Ribeiro, Marcelo Rubinho, Daiane B. Seriacopi, Fernando Silva, Rodolfo Valentim and Aline Vidotto.

MJFV, MH and YAA acknowledge funding from the European Research Council (ERC) under the European Union's Horizon 2020 research and innovation programme (grant agreement No 772663). The authors acknowledge Interstellar Institute's program "With Two Eyes" and the Paris-Saclay University's Institut Pascal for hosting discussions that nourished the development of the ideas behind this work. This research has used data, tools or materials developed as part of the EXPLORE project that has received funding from the European Union’s Horizon 2020 research and innovation programme under grant agreement No 101004214.

AMM's work and Optical/NIR Polarimetry at IAG has been supported over the years by several grants from São Paulo state funding agency FAPESP, especially 01/12589-1 and 10/19694-4. AMM has also been partially supported by Brazilian agency CNPq (grant 310506/2015-8). AMM graduate students have been provided grants over the years from Brazilian agency CAPES.

CVR thanks Conselho Nacional de Desenvolvimento Científico e Tecnológico - CNPq (Proc: 310930/2021-9).
\end{acknowledgments}

\facilities{LNA:BC0.6m}

\bibliography{refs}{}
\bibliographystyle{aasjournal}



\end{document}